\documentclass[epj]{svjour}
\usepackage{graphics}
\usepackage{floatflt}
\usepackage{color}
\usepackage{colortbl}
\newcommand{\Pt}{{P_t}}
\newcommand{\dphi}{\Delta\phi}
\newcommand{\phigj}{\phi_{(\gamma,jet)}}

\newcommand{\gpj}{~``$\gamma+jet$''~}
\newcommand{\rrr}{\to} 
\newcommand{\pth}{\hat{p}_{\perp}^{\;min}}

\newcommand{\Ptg}{\Pt^{\gamma}}
\newcommand{\tg}{\tilde{\gamma}}
\newcommand{\Pttg}{\Pt^{\tg}}
\newcommand{\QQ}{Q^{~2}}
\newcommand{\lt}{\!\!<\!\!}

\newcommand{\GC}{GeV/c}

\newcommand{\Gvc}{\footnotesize{$(GeV/c)$} }
\newcommand{\com}{qg\to q+\gamma}
\newcommand{\ann}{q\bar{q} \to \gamma + g}
\newcommand{\hmm}{\hspace*{-1.3mm}}      
\newcommand{\coltab}{0.95}

\def\baselinestretch{1.7}

\begin{document}
\title{On the possibility of measuring the gluon distribution in proton with
``$\gamma+jet$'' events at LHC
}
\titlerunning{On the possibility of measuring the gluon distribution at LHC}
\authorrunning{D.V.~Bandurin, N.B.~Skachkov}
\author{D.V.~Bandurin\inst{1} \and N.B.~Skachkov\inst{1}
\thanks{\emph{Present address:} Joliot-Curie 6, JINR, 141980, Dubna, Moscow region, Russia}%
}                     
%
%
\institute{Joint Institute for Nuclear Research, Dubna, Russia}
\date{Received: ~~~~~~~~~~~ / Revised version: ~~~~~~~~~~~}
%
\abstract{
The numbers of the \gpj events suitable for a determination of the gluon distribution function
$f^p_g(x,Q^2)$ in a proton at the LHC for various intervals of $x$ and $Q^2$
are estimated. The contributions of different background sources 
are studied. The values of discrimination powers between 
quark and gluon jets as well as between a single photon and the products of $\pi^0, \eta, \omega$ and 
$K^0_s$ mesons decaying through the neutral channels are applied to estimate the final contributions 
of different event types to the \gpj production in various intervals of $x$ and $Q^2$. 
The PYTHIA event generator was used to produce physical events for this analysis.
} 

\maketitle

\onecolumn
\section{Introduction.}
\label{intro}

The modeling of the production processes of many new particles (Higgs boson, SUSY particles)
in the forthcoming LHC experiments as well as future physical analysis of corresponding
measurement are heavily based on the knowledge of 
gluon distribution  in a proton $f^p_g(x,Q^2)$ \cite{Higgs}
\footnote{For example, the production of Standard Model Higgs boson is mainly caused by
gluon-gluon fusion $gg\to H$ over the entire mass range \cite{Higgs}.}.
For this reason the study of the possibility of 
the measuring gluon density  directly in
the LHC experiments (especially in the kinematic region of small $x$ and high $Q^2$) is of a big interest.

One of the promising channels for this measurement 
is an inclusive prompt photon production \cite{Au_Aa}
\begin{eqnarray}
pp\rightarrow \gamma^{dir} + X.
\label{eq:incl}
\end{eqnarray}
The region of photon transverse momentum $\Ptg$, 
reached by UA1 \cite{UA1}, UA2 \cite{UA2}, CDF \cite{CDF1} and 
D0 \cite{D0_1} experiments,
extends up to $\Ptg \approx 60~ GeV/c$ and, according to recent
results \cite{D0_2}, up to $\Ptg \approx 105~GeV/c$. 
These data together with the later ones
(see references in \cite{Au0}--\cite{Fr1}) and E706 \cite{E706}, UA6 \cite{UA6} results give, in principle, 
an opportunity for tuning the form of gluon distribution 
(see \cite{Au2,Vo1,Mar,CTEQ}). The rates and an estimation for cross sections of inclusive
photon production at LHC are given in \cite{Au_Aa} (see also \cite{AFF}). 

Here we consider the process of a direct photon production in association with 
an opposite-side jet
\cite{Cont,Au2} (for experimental results see \cite{ISR,UA2_g,CDF2})\\[-33pt]
\begin{eqnarray}
pp\rightarrow \gamma^{dir} + jet + X.
\label{eq:gpj}
\end{eqnarray}

In QCD leading order the processes (\ref{eq:incl}) and (\ref{eq:gpj}) are caused 
by two subprocesses
\footnote{A contribution of another possible NLO channel $gg\rrr g\gamma$
was found to be still negligible even at LHC energies.}:
the ``Compton-like" scattering \\[-22pt]
\begin{eqnarray}
\hspace*{0.14cm} qg\to q+\gamma 
\label{eq:gluon} 
\end{eqnarray}
\vskip-4mm
\noindent
and the ``annihilation'' subprocess\\[-22pt]
\begin{eqnarray}
\hspace*{0.12cm} q\overline{q}\to g+\gamma.  
\label{eq:annih} 
\end{eqnarray}
\vskip-2mm

\noindent
The first one gives a dominant contribution to the cross sections of 
(\ref{eq:incl}) and (\ref{eq:gpj}) \cite{Au0,Cont,Au2} and serves as
``signal'' subprocess due to its direct connection with the gluon distribution.

The study of ``$\gamma + jet$'' process (\ref{eq:gpj}) is 
a more preferable as compared
with the inclusive direct photon production process (\ref{eq:incl})
from the viewpoint of extraction of information on the gluon distribution $f^p_g(x,Q^2)$
\footnote{A detailed study of \gpj events and different aspects of their application
can be found in \cite{CMSthick,BALD00}.}.
First of all, it is explained by a higher value of a purity of the process (\ref{eq:gpj}), 
for which the signal-to-background ($S/B$) ratios are several times higher than 
$S/B$ ratios to process (\ref{eq:incl}) \cite{P5,PartIII}.
Secondly, while the cross section for the process (\ref{eq:incl}) is given as an integral 
over the parton distribution functions (PDF) of a proton $f^p_a(x,Q^2)$,   
the  cross section of the process (\ref{eq:gpj}) is expressed directly 
(at $\Ptg \geq 30/~ GeV/c$
\footnote{i.e. in the region where ``$k_T$ smearing effects'' should not be important 
(for example, see \cite{Hu2}).})
through these PDFs: 
~\\[-20pt]
\begin{eqnarray}
\frac{d\sigma}{d\eta_1d\eta_2 d(\Pt^{\!\!\gamma})^2} = \sum\limits_{a,b}\,x_a\,f^p_a(x_a,Q^2)\,
x_b\,f^p_b(x_b,Q^2)\frac{d\sigma}{d\hat{t}}(a\,b\rightarrow 1\,2),
\label{eq:cross}
\end{eqnarray}
\noindent
where $a,b=q,\bar{q},g$; $1,2=q,\bar{q},g,\gamma$. 
The incident parton momentum fractions $x_a, x_b$ may be reconstructed from the final
state photon and jet pseudorapidities 
$\eta_1=\eta^\gamma$, $\eta_2=\eta^{jet}$ and $\Ptg$ according to formula
\footnote{see, for instance, \cite{Owe,Cont}.}
~\\[-20pt]
\begin{eqnarray}
x_{a,b} \,=\,\Ptg/\sqrt{s}\cdot \,(exp(\pm \eta_{1})\,+\,exp(\pm \eta_{2})).
\label{eq:x_gl}
\end{eqnarray}


Thus, formula (\ref{eq:cross}) with the knowledge of the experimentally determined
triple cross section in the intervals of $\Delta\eta^\gamma$, $\Delta\eta^{jet}$ and
$(\Pt^{\!\!\gamma})^2$ and with account of
results of independent measurements of $q, \,\bar{q}$ distributions \cite{MD1} 
allows the gluon distribution $f^p_g(x,Q^2)$ to be determined 
after  essential reduction of a background contribution.


The $\Ptg$ distributions of the number of signal events remained after application of strict
selection criteria, proposed in \cite{P5,P1}, were presented earlier
in \cite{BALD00,DIS03,GLU,DMS}
\footnote{Analogous estimations for the Tevatron were done in 
\cite{D0_Note,BALD02}.}.
Those selection criteria allow to reduce considerably the background 
to ``$\gamma^{dir}+jet$'' process 
(\ref{eq:gpj}) and select the events with a suppressed initial state radiation.
Here we  present the detailed study of background fraction in different
$x$- and $Q^2$- intervals and show that ``gluonic'' subprocess (\ref{eq:gluon}) gives
a noticeable contribution \cite{DIS03}.

This paper is organized as follows. The main background sources  are discussed in section 2.
In section 3 we list the selection criteria used to enhance the content of
the signal events (\ref{eq:gluon}) in the selected data sample.
In section 4 the numbers of the events suitable for an extraction 
of the gluon distribution function $f^p_g(x,Q^2)$ are estimated. The contribution of events of other types
in different $x$- and $Q^2$- intervals are also shown in this section.
A possibility of the further background events suppression 
by an account of the discrimination efficiencies between
a single photon and $\pi^0, \eta, \omega$ and $K^0_s$ mesons 
as well as between quark and gluon jets is also demonstrated.

\section{Background sources.}
%

The background to the events based on the process (\ref{eq:gpj})
is mainly caused by: 

--- the events with high $\Pt$ photons produced in the neutral decay channels 
of $\pi^0, \eta, \omega$ and $K^0_s$ mesons 
\footnote{As it was shown in \cite{GMO} the charged decay channels of those mesons
can be strongly suppressed even without a tracker information.}. 
%
%
%

%
--- the events with the photons radiated from a quark (i.e. bremsstrahlung 
photons) in the next-to-leading order QCD subprocesses of the $qg\to qg$, $qq\to qq$ and 
$q\bar{q}\to q\bar{q}$ scattering \cite{P5,PartIII}.

The background events of the first type  will be called below as the 
``$\gamma\!-\!mes$'' events while the events of the second type as the ``$\gamma\!-\!brem$'' ones.
A more detailed information about fundamental QCD subprocesses from which originate
``$\gamma\!-\!mes$'' and ``$\gamma\!-\!brem$'' events  is presented in section 3.

The background may be also caused by ``$e^{\pm}$ events'' which contain one jet and
$e^{\pm}$ as a direct photon candidate. The value of the fraction of these events in the total
background was estimated in \cite{P5,PartIII} (see also sections 3, 4).


The background containing the ``$\gamma\!-\!mes$'' events 
can be significantly suppressed by the event selection criteria, pointed 
in \cite{P5,PartIII,P1}. It may be achieved, first of all, due to very strict photon isolation criteria 
because a parent meson ($\pi^0, \eta, \omega$ or $K^0_s$) is usually surrounded by other particles.
Additional rejection factors were obtained from a full GEANT simulation of the physical processes
in the CMS detector \cite{CMS_EC} 
where for the Barrel region ($|\eta|\!<\!1.4$) we used  an information 
from the electromagnetic calorimeter (ECAL) cells only
\footnote{A preshower detector is not foreseen currently in the Barrel region of the CMS
detector \cite{CMS_EC}.}
\cite{BS_MES} while for the Endcap region 
($1.4\!<\!|\eta|\!<\!2.5$) the results of the analysis of hits in the preshower detector 
\cite{BarnPS} were applied.

An especial attention should be paid to the events containing the bremsstrahlung photons. 
They are also noticeably rejected by the selection cuts but still
constitute a significant part of the total background 
\cite{P5,PartIII}. 

The simulation, performed with a help of the Monte Carlo event generator PYTHIA \cite{PYT}, 
has shown that in the selected ``$\gamma+jet$'' event samples
the most part of ``$\gamma\!-\!brem$'' events contain a gluon jet (see section 4).
For this part of events, 
%
as well as for a part of ``$\gamma\!-\!mes$'' events with  a gluon jet,
%
one can take into consideration the quark/gluon separation efficiencies found earlier 
in \cite{BS_QG}
\footnote{see also \cite{GL_TR,NN_IJ}.}.
The number of remained background events also must  be well estimated
in order to separate their contribution from one of the ``$\gamma^{dir}+jet$'' events (\ref{eq:gpj}).

\section{Definition of selection cuts~~.}
{\vskip-33pt
\hspace*{53.5mm} ~\footnote{In this section we follow mostly the selection criteria from 
\cite{P5,P1}.}}                                                                    

\vskip10pt

1. Only the events with one jet and one ``$\gamma^{dir}$-candidate''
(in what follows we shall denote it also as $\tg$ and call the
``photon'' for brevity) with\\[-15pt]
\begin{equation}
\Pt^{jet}\geq 30 \;GeV/c \quad {\rm and} \quad \Pt^{\tg} \geq 40~ GeV/c~.
\label{eq:sc1}
\end{equation}
\noindent
are considered here. 
In the simulation the signal (most energetic $\gamma$ or $e^\pm$ together 
with surrounding particles) is considered as a candidate for a direct photon
if it fits into one CMS calorimeter tower having the size of $0.087\times0.087$ in 
the $\eta-\phi$ space \cite{CMS_EC}.

%

For all applications a jet is defined according to the PYTHIA
jetfinding algorithm LUCELL \cite{PYT}.
%
%
In the $\eta-\phi$ space the jet cone of radius $R$  counted from the jet initiator cell is
taken to be $R\!=\!((\Delta\eta)^2+(\Delta\phi)^2)^{1/2}\!=\!0.7$.

\noindent
2. To suppress the contribution of background processes, i.e. to select mostly the events 
with the ``isolated'' photons and to discard the events that fake a direct photon signal, 
we restrict:

a) the value of the scalar sum of $\Pt$ of hadrons and other particles surrounding
a photon within a cone of $R^{\gamma}_{isol}\!=\!( (\Delta\eta)^2+(\Delta\phi)^2)^{1/2}\!=\!0.7$
(``absolute isolation cut")\\[-15pt]
\begin{equation}
\sum\limits_{i \in R_{isol}} \Pt^i \equiv \Pt^{isol} \leq \Pt_{CUT}^{isol};
\label{eq:sc2}
\end{equation}
\vspace{-2.6mm}

b) the value of a fraction (``fractional isolation cut'')\\[-15pt]
\begin{equation}
\sum\limits_{i \in R_{isol}} \Pt^i/\Pt^{\tg} \equiv \epsilon^{\gamma} \leq
\epsilon^{\gamma}_{CUT}.
\label{eq:sc3}
\end{equation}

\noindent
3. Only the events having no tracks
\footnote{i.e. charged particles as we use the PYTHIA level of simulation.}
with $\Pt>1~GeV/c$ contained inside the cone of $R=0.4$ around a $\gamma^{dir}$-candidate are accepted.

\noindent
4. 
To suppress the background events with photons resulting from
$\pi^0$, $\eta$, $\omega$ and $K_S^0$ meson decays, we require the absence of a high $\Pt$ hadron
in the tower containing the $\gamma^{dir}$-candidate:\\[-1pt]
\begin{equation}
\Pt^{hadr} \leq 7~ GeV/c.
\label{eq:sc5}
\end{equation}

\noindent
At the PYTHIA level of simulation this cut may effectively takes into account 
the imposing of an upper cut on the energy deposited in the cells of hadronic calorimeter (HCAL)
that are behind the ECAL signal cells fired by the photon. 
In real experimental conditions one can require for a fraction of the photon energy, 
deposited in ECAL to be greater than some threshold
\footnote{e.g. to be greater than $0.95$ (as it was used at D0 \cite{D0_2})}.

\noindent  
5. We select the events with the vector $\vec{\Pt}^{jet}$ being ``back-to-back" to
the vector $\vec{\Pt}^{\tg}$ (in the plane transverse to the beam line)
within the azimuthal angle interval $\dphi$ defined by the equation:\\[-20pt]
\begin{equation}
\phigj=180^\circ \pm \Delta\phi.
\label{eq:sc7}
\end{equation}
~\\[-9pt]
The angle $\phigj$ between $\vec{\Pt}^{\tg}$ and $\vec{\Pt}^{jet}$ vectors  is
calculated from  the expression
~$\vec{\Pt}^{\tg}\vec{\Pt}^{jet}=\Pt^{\tg}\Pt^{jet} cos(\phigj)$ ~
with ~$\Pt^{\tg}=|\vec{\Pt}^{\tg}|$ and $\Pt^{jet}=|\vec{\Pt}^{jet}|$.
The value of $\Delta\phi$ may be chosen from the interval $5^\circ\div 15^\circ$  
for various energies.

\noindent
6. 
We also choose only the events that do not have any other, except one jet,
minijet (or cluster) high $\Pt$ activity with the $\Pt^{clust}$
higher than some threshold $\Pt^{clust}_{CUT}$ value. Thus,
we select events with\\[-14pt]
\begin{equation}
\Pt^{clust} \leq \Pt^{clust}_{CUT},
\label{eq:sc8}
\end{equation}
where clusters are found by the same jetfinder LUCELL used to determine the main jet 
in the event. The most effective restrictions are $\Pt^{clust}_{CUT}=5\div 15 ~GeV/c$.
Their choice will be caused mostly by the gained statistics and $\Pttg$ value (for higher
$\Pttg$ a weaker $\Pt^{clust}_{CUT}$ can be used).

\noindent
7. The  events containing $e^\pm$ as a photon candidate are mainly caused by
the  subprocesses ~$q\,g \to q' + W^{\pm}$~ and ~$q\bar{~q'} \to g + W^{\pm}$~  
with the subsequent decay $W^{\pm} \to e^{\pm}\nu$.
To reduce a contribution from 
these events \cite{P5,PartIII} we shall
select only events having a small value of missing transverse momentum 
$\Pt^{miss}$. So, we also use the following cut:\\[-24pt]
\begin{eqnarray}
\Pt^{miss}~\leq \Pt^{miss}_{CUT}.
\label{eq:sc11}
\end{eqnarray}

Finally, in what follows we shall set the values of the cut parameters
(besides those pointed above explicitly) as specified below:
~\\[-14pt]
\begin{eqnarray}
\Pt^{isol}_{CUT}=2\;GeV/c,\;\; 
{\epsilon}^{\gamma}_{CUT}=5\%, \;\;
\dphi \leq 15^{\circ}, \;
\nonumber
\end{eqnarray}
~\\[-16mm]
\begin{eqnarray}
\Pt^{clust}_{CUT}=10\;GeV/c,\;\;
\Pt^{miss}_{CUT}=10\;GeV/c. 
\label{eq:sc}
\end{eqnarray}


%
\section{Determining the numbers of events and reducing the background.}
%

To estimate a background to the signal events we have done a simulation
using the Monte Carlo event generator PYTHIA with
a mixture of all existing in PYTHIA QCD and SM subprocesses with large cross sections 
\footnote{They have ISUB=11--20, 28--31, 53, 68 according to the process numbers in PYTHIA \cite{PYT}.},
including subprocesses (\ref{eq:gluon}) and (\ref{eq:annih}) 
\footnote{with ISUB=14 and 29  in notations of PYTHIA \cite{PYT}.}.

The total cross section of the background  subprocesses exceeds  the cross section 
of the subprocesses (\ref{eq:gluon}) and (\ref{eq:annih})
by more than three orders of magnitude.
%
%
The GRV 94L parameterization of the parton distribution functions is used as a default one.

Five generations (each of about 60--90 million events)
with different values of minimal transverse momentum of a hard subprocess 
\footnote{CKIN(3) parameter in PYTHIA.}
$\pth$ were done: $\pth$ = 40, 70, 100, 140 and 200 $GeV/c$.
The cross sections of the abovementioned subprocesses define the rates of corresponding physical
events and, thus, appear here as weight factors.
The selection criteria of section 3 were applied then to the generated events.

The total numbers of these events, i.e. events originated 
from subprocesses (\ref{eq:gluon}) and (\ref{eq:annih})  as well as ``$\gamma\!-\!brem$'' 
and ``$\gamma\!-\!mes$'' events, are  presented (being divided by the factor of $10^3$)
in Table \ref{tab:S+B_0} 
for each $x$ and $\QQ$ interval ($\QQ\equiv (\Ptg)^2$) for the integrated luminosity 
\footnote{This value is intended to be accumulated during one year of LHC running at luminosity
$L=10^{33}~cm^{-2} s^{-1}$.}
$L_{int}=10~fb^{-1}$. 
The momentum fractions $x_a$ and $x_b$ 
of the initial state partons were calculated via the photon and jet parameters
according to formula (\ref{eq:x_gl}) \cite{Owe,Cont}.
The right-hand columns of this table shows, for a convenience, the correspondence
of $\QQ$ interval to the $\Ptg$ interval.

One can  see from Table \ref{tab:S+B_0} that at $40<\Ptg<50~GeV/c$ 
the total number of events is about 10 million and it drops
to 24~200 at $200<\Ptg<283~GeV/c$, i.e. with five-fold increase of $\Pt^{\tg}$
the spectrum drops about 400 times.

The contribution from the background ``$e^\pm$ events'' was not included in Table \ref{tab:S+B_0}. 
The number of these type events was estimated in 
\cite{P5,PartIII}
\footnote{It was found that 
after  application of the selection criteria from section 3 and taking
a track finding efficiency to be equal to $85\%$ (being averaged over all
pseudorapidity range) \cite{CMS_TRA}
a contribution of the $e^\pm$ events (having the isolated $e^\pm$ with $\Pt^{e}>40~GeV/c$)
to the total background reduces to less than $1\%$ at
$40\leq\Pt^e \leq 70 ~GeV/c$ and to about $5\%$ at $\Pt^e \geq 100 ~GeV/c$.}
and found to be very small as compared with other background types.
Thus, in what follows we shall concentrate on more sizable background.

Now let us look at the contributions of different event types in various $x$ and $\QQ$ intervals.
The events selected after passing the criteria of section 3 were classified 
in accordance with the origin of the produced $\gamma^{dir}$-candidates. 
So, we consider separately those that contain the direct photons 
(produced in subprocesses (\ref{eq:gluon}) and (\ref{eq:annih})) and
those that have $\gamma^{dir}$-candidates appearing due to the radiation from quarks 
(``$\gamma\!-\!brem$'' events) 
or from the $\pi^0$, $\eta$, $\omega$ and $K_s^0$ meson decays (``$\gamma\!-\!mes$'' events).
All these contributions are presented in Tables 1A--4A of Appendix 
in a form of number of events divided by the factor of $10^3$.
The numbers of the events based on the Compton (\ref{eq:gluon}) and annihilation 
(\ref{eq:annih}) subprocesses are shown in Tables 1A and 2A while
the numbers of the ``$\gamma\!-\!brem$'' 
and ``$\gamma\!-\!mes$'' events can be found in Tables 3A and 4A, respectively
\footnote{See also \cite{P5,PartIII} for a more detailed information about background composition.}.
These numbers were obtained after passing the selection criteria of section 3. 
The fractions of each event type, calculated for
a given interval of $\Pttg$, are presented in Fig.~\ref{fig:proc}a
($100\%$ is taken for all types of events).

We see that the main part of the background is due to ``$\gamma\!-\!brem$'' events 
and the combined contribution of ``$\gamma\!-\!brem$'' and ``$\gamma\!-\!mes$'' events
into the total number of events
varies from about $23\%$ at $40<\Pttg<50~GeV/c$ to about $6\%$ at $100<\Pttg<140~GeV/c$
and drops to $4\%$ at $200<\Pttg<283~GeV/c$.

We would like to stress that the essential point of our analysis is the
study of the background contributions 
after application of the cuts for selecting the \gpj events with
 a limited cluster/minijet activity and a clean $\gamma-jet$ topology
\footnote{See \cite{P5,PartIII} where 
the dynamics of application of the selection criteria is demonstrated.}.
Only in this case a contribution of ``$\gamma\!-\!brem$'' and ``$\gamma\!-\!mes$''
events can be decreased noticeably
\footnote{It was shown in \cite{P5,PartIII} 
that, for instance, at $\Pt^{\gamma}>100 ~GeV/c$, the application of 
``photonic'' cuts, usually used to select inclusive photon ``$\gamma+X$'' events,
gives $S/B=1.9$ only, while a further
account of ``hadronic'' and topological cuts for selection of \gpj events leads to $S/B=17.6$,
i.e {\it to the increase of $S/B$ by about one order of magnitude} 
(here $S$ is a total contribution from the events based on the subprocesses (\ref{eq:gluon}) and 
(\ref{eq:annih}) and $B$ is a contribution from the sum of ``$\gamma\!-\!brem$'' and 
``$\gamma\!-\!mes$'' events). The application of other cuts that limit a $\Pt$ activity out of 
the \gpj system may lead to the following $20-30\%$ increase of the $S/B$ ratio 
\cite{P5,PartIII}.}.

The selection criteria of section 3 are not final and are moderate enough. 
The results of their application may change if we shall 
vary some cuts. So, for example, a stronger limitation of cluster activity (\ref{eq:sc8})
by $\Pt^{clust}_{CUT}=5~GeV/c$ would lead to a further substantial decreasing of the numbers of 
``$\gamma\!-\!brem$'' and ``$\gamma\!-\!mes$'' events \cite{P5,PartIII}.

The contribution of ``$\gamma\!-\!mes$'' events can be also reduced by the account of the difference
between a single photon and the $\pi^0$, $\eta$, $\omega$ and $K_s^0$ meson 
signals produced in the detector.

To take into account the discrimination efficiencies between a single photon and the photons produced
via multiphoton decays of $\pi^0, \eta$, $\omega$ and $K^0_s$ mesons ($\epsilon^{\gamma/mes}$), 
the results of papers \cite{BS_MES} and \cite{BarnPS} were used. 
The efficiencies found in \cite{BS_MES} were obtained by the analysis
of the ECAL crystal cells only in the Barrel region ($|\eta|\!<\!1.4$) while the efficiencies in
\cite{BarnPS} are found from the analysis of hits in the preshower detector in the Endcap region
($1.4\!<\!|\eta|\!<\!2.5$). 
The results of \cite{BS_MES} and  \cite{BarnPS} are briefly 
the following:  the rejection efficiency of the neutral pion is about
$49-67\%$ (depending on an energy) in the Barrel region and it ranges as
$45-71\%$ for the Endcap region.
The single photon selection efficiencies ($\epsilon^{\gamma}_{sel}$) 
were set to $70\%$ and $91\%$ in the first and second cases respectively
\footnote{With the same $\epsilon^{\gamma}_{sel}=70\%$ one can also reject about $90-95\%$ 
of ``$\eta-$meson events'' and $55-92\%$ of ``$K^0_s-$meson events'' \cite{BS_MES} 
in the Barrel region. For the Endcap 
the respective rejection efficiencies were taken here equal to those obtained for $\pi^0$ meson.
At the same time it is worth to note that the main contribution to the ``$\gamma-mes$'' background
comes from the ``$\pi^0$-events'' ($\sim 62-65\%$ in the interval $40\lt\Pttg\lt 140~GeV/c$) \cite{P5,PartIII}.
}.

The results of applications of the described above $\gamma/$meson separation
efficiencies to the ``$\gamma-mes$'' events themselves are placed in Table 8A (compare with
Table 4A) and in Tables 5A--7A of Appendix for other event types.
%
%
Thus, we see that for the ``$\gamma-mes$'' events the reduction factor of about $2-3$ for
$40<\Pttg<100~GeV/c$ can be obtained
with a loss of $16-19\%$ of events of other types with a single photon in the final state.
The total numbers of all events left after account of the $\epsilon^{\gamma/mes}$ separation
efficiencies are presented in Table \ref{tab:S+B_1}.


The physical models implemented in PYTHIA allows to get an idea about a possible 
origination of the ``$\gamma-brem$'' and ``$\gamma-mes$'' events.
Tables \ref{tab:bg_or_gr} and \ref{tab:bg_or_ms} show  the relative
contributions of four main (having the largest cross sections) fundamental QCD subprocesses 
$qg\to qg$, $qq\to qq$, $gg\to q\bar{q}$ and $gg\to gg$ into a production of 
the ``$\gamma-brem$'' and ``$\gamma-mes$'' events 
selected with the criteria 1--7 of section 3 for three $\Pttg$ intervals
\footnote{The sum over contributions
from the four considered QCD subprocesses in some lines of Tables \ref{tab:bg_or_gr} and 
\ref{tab:bg_or_ms} is less than 100$\%$. The remained percentages correspond to
other subprocesses (like $q\bar{q}\to q\bar{q}$ or $qg\to q^\prime W^{\pm}$).
The errors in those tables are statistical and caused by the number of entries for various 
background types after application of the criteria 1--7 of section 3.}.
One can see from these tables
that most of ``$\gamma-brem$'' and ``$\gamma-mes$'' events (from $70$ to $80\%$) 
still originate from ``gluonic''
$qg\to qg$, $gg\to q\bar{q}$ and $gg\to gg$ subprocesses 
with dominant contribution from the first subprocess.

The analysis of the PYTHIA simulation output also shows that practically in all of selected 
``$\gamma\!-\!brem$'' events
the ``bremsstrahlung photons'' are produced in the final state of the fundamental subprocess.
They are radiated from the outgoing quarks in the case of the first three subprocesses
or can appear as the result of string breaking in the case of $gg\to gg$ scattering. The last
mechanism, naturally, gives a small contribution into the ``$\gamma\!-\!brem$'' events production.
In the first case the selected (see section 3)
photon carries away almost all energy of a quark in the final state. 
The events of this kind have mostly a gluon jet ($70.6\%$ of events for $40<\Pttg<71 ~GeV/c$
interval and $58.7\%$ of events for $141<\Pttg<283 ~GeV/c$) with the photon radiated 
in back-to-back direction to the jet in $\phi$ plane. In the second case ($gg\to gg$ based events) 
a remained jet is practically always of the gluon type.

As for ``$\gamma\!-\!mes$'' events, it is naturally to expect
that in the events based on the $qg\to qg$ scattering after suppression of the cluster 
activity by the cut $\Pt^{clust}\!\!<\! 10 ~GeV/c$ (see (\ref{eq:sc8}))
a remained jet can originate with an equal probability
from a quark as well as from a gluon (50$\%$ by 50$\%$) while in the events based on the
$qq\to qq$, $gg\to q\bar{q}$ ($gg\to gg$) subprocesses the jet is always of the quark (gluon) type.

Thus, one can conclude that about $73\%$~($40\%$), $70\%$~($36\%$) and $59\%$~($33\%$) 
of the ``$\gamma\!-\!brem$'' (``$\gamma\!-\!mes$'') events have a gluon jet in the selected 
one-jet events in $\Pttg$ intervals $40\div 71$, $71\div 141$ and $141\div 283 ~GeV/c$, respectively. 
%
%

For the following suppression of the contributions from
``$\gamma\!-\!brem$'' and ``$\gamma\!-\!mes$'' events 
having a gluon jet in the final state one can apply
the quark/gluon separation efficiencies ($\epsilon^{q/g}$) obtained earlier in \cite{BS_QG}.
The results of \cite{BS_QG} shows that with account of the  quark jet
selection efficiency of about $65-67\%$ it is possible to reject $73-81\%$ of gluons jets
\footnote{This efficiency slightly depends on the jet transverse momentum $\Pt^{jet}$ and 
pseudorapidity $\eta^{jet}$ \cite{BS_QG}.}
for $\Pt^{jet}$ varying from $40$ to $200~GeV/c$.

The numbers of different types of events after an account of the both $\epsilon^{\gamma/mes}$ 
and $\epsilon^{q/g}$ separation efficiencies  are presented 
in Tables 9A--12A of Appendix while their fractions (in $\%$)
are shown in Fig.~1c.

By comparing Tables 7A and 11A one can see that the numbers of ``$\gamma\!-\!brem$'' events
\footnote{that are, in fact, irreducible by using only photon information after
application of the strong isolation cuts (\ref{eq:sc2}) and (\ref{eq:sc3}).}
are reduced in $2.5-3$ times at the cost of $35\%$ loss of the events based on 
subprocess (\ref{eq:gluon}) and their fraction in the total number of events
becomes about $8\%$ at $40<\Pttg<50~GeV/c$ and about $2\%$ at $140<\Pttg<200~GeV/c$.
{\it The total contributions  of the ``$\gamma\!-\!brem$'' and ``$\gamma\!-\!mes$'' events 
in the same $\Pttg$ intervals compose $13.2\%$ 
and $2.8\%$, respectively}.

%
We see also that the account of $\epsilon^{q/g}$ separation efficiency
reduces a contribution of the events originated from annihilation subprocess (\ref{eq:annih}) 
to the total number of events
(especially at higher $\Ptg$) to the size of about $3-5\%$ over all considered $\Ptg$ range
(see Fig.~1 and Tables 1A--8A).
%

The final numbers of all \gpj events for the luminosity $L_{int}=10 ~fb^{-1}$ at 
different $x$ and $Q^2$ intervals after an account of the both separation efficiencies
are given in Table \ref{tab:S+B_2}. 
We see that after passing all selection cuts and application of the $\epsilon^{\,\gamma/mes}$ 
and $\epsilon^{\,q/g}$ efficiencies one can get about 5 million events at 
the $40<\Pttg< 50 ~\GC$ interval, about 200~000 at $100<\Pttg<141 ~\GC$ and 
about 11~000  at the last considered interval $200<\Pttg<283 ~\GC$.
{\it The total expected statistics on the \gpj events, left after account of
the $\epsilon^{\,\gamma/mes}$ and $\epsilon^{\,q/g}$ efficiencies,
is about $9\cdot 10^6$ events}. The final contributions
of different subprocesses in various $x$ and $Q^2$ intervals are presented in 
Tables 9A--12A.

\section{Conclusion.}

\begin{figure}
\resizebox{0.88\textwidth}{!}{
   \hspace*{40mm} \includegraphics{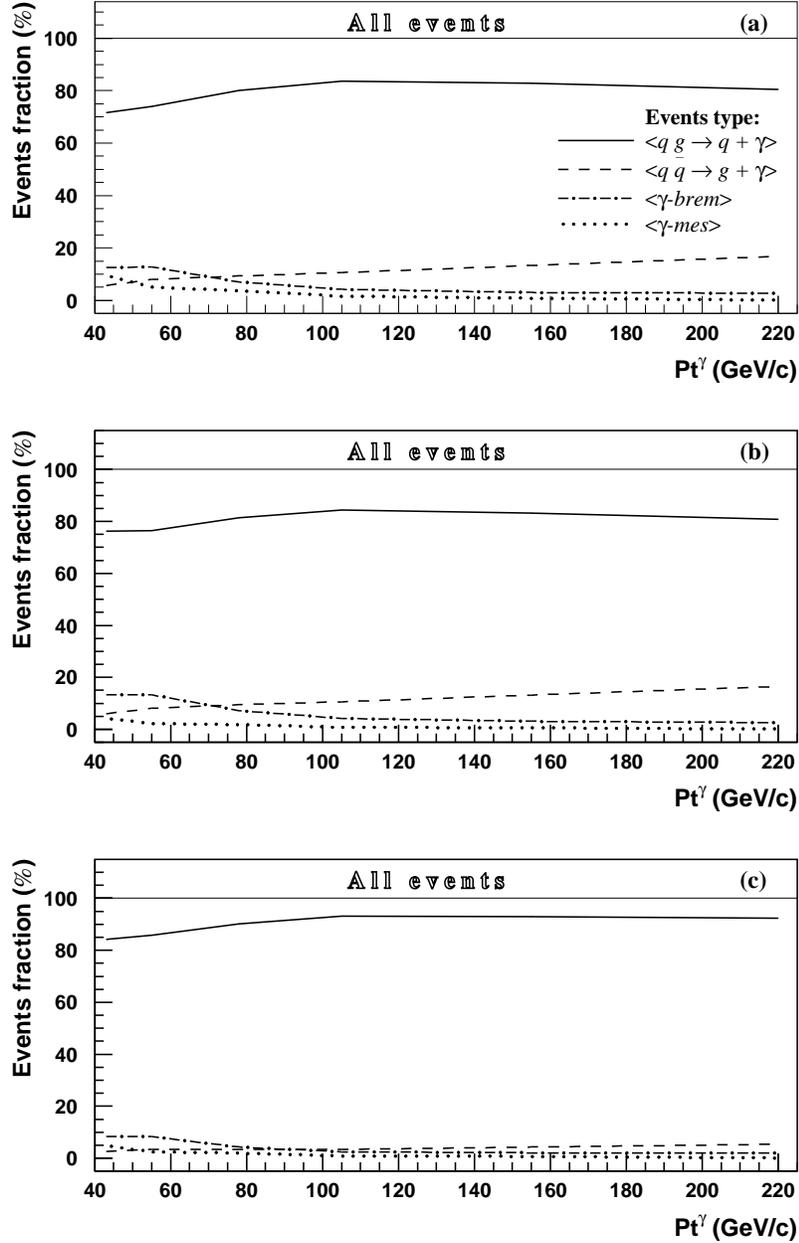}
}
\caption{The contributions of various events types to the total number of events as a function 
of $\Pttg$ presented for three cases: {\bf(a)}~ No separation efficiency is taken into account, 
{\bf(b)}~ $\epsilon^{\,\gamma/mes}$ separation efficiencies are taken into account 
and {\bf(c)}~ $\epsilon^{\,\gamma/mes}$ and $\epsilon^{\,q/g}$ separation
efficiencies  are taken into account.}
\label{fig:proc}
\end{figure}

The results presented above show that during one year of the LHC running at low luminosity
($L=10^{33} \,cm^{-2}s^{-1}$) one can collect
after application of the proposed selection criteria
the clean sample of ``$\gamma^{dir}+jet$'' events with
a sufficient statistics to determine a
gluon density in a proton in new  kinematic region of $2\cdot 10^{-4}\leq x \leq 1$ and
$1.6\cdot10^3\leq Q^2\leq2\cdot10^5 ~(GeV/c)^2$. 
%
%
At the same time the combined contribution of 
``$\gamma-brem$'' events and 
``$\gamma-mes$'' events is estimated to be about $23\%$ at $40<\Pttg<50~GeV/c$ and it drops
to $4\%$ at $200<\Pttg<283~GeV/c$ (see Tables 1A--4A and Fig.~\ref{fig:proc}a).

The given estimations on 
contributions of the ``$\gamma-brem$'' and ``$\gamma-mes$'' events are not final yet. 
For instance, a stronger limitation $\Pt^{clust}_{CUT}=5~GeV/c$ 
(see (\ref{eq:sc8})) would lead to a following substantial (about $30\%$) 
reduction of their contribution \cite{P5,PartIII}.

With an additional account of discrimination efficiencies between
single photons and $\pi^0, \eta, K^0_s$ mesons 
as well as  those between quark and gluon jets \cite{BS_MES,BarnPS,BS_QG}
one can increase noticeably the purity of the selected samples of
the ``$\gamma^{dir}+jet$'' events (see Tables 9A--12A of Appendix and Fig.~1). 
A possibility to obtain better background rejection factors 
will depend on the chosen values of single photon and
quark jet selection efficiencies 
\footnote{Let us remind that 
the single photon selection efficiencies equal to $70\%$ and $91\%$ for the Barrel and Endcap 
regions and quark jet selection efficiencies  equal to about $65\%$ were chosen here for 
the given estimations.}
which are in their turn will be caused by a gained statistics of the \gpj events.

It is also worth mentioning that a full simulation
\footnote{We mean a full simulation of the detector response with the following digitization
and reconstruction of signals from physical objects.}
of the signal and background processes is rather difficult 
due to a very small selection efficiency for the background events
($\approx 0.01-0.05\%$ depending on an energy) \cite{P5,PartIII} what, in its turn, 
requires huge computational resources to collect the background events statistics sufficient 
for the analysis.

\begin{figure}
\resizebox{0.85\textwidth}{!}{
    \hspace*{70mm}\includegraphics{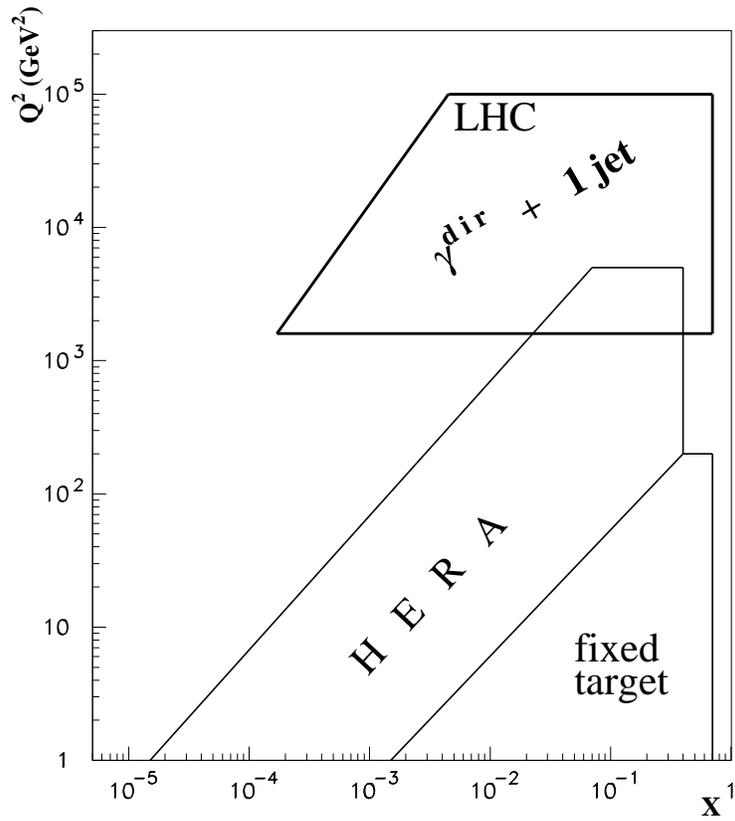}
}
\caption{LHC  $(x,Q^2)$ kinematic region for $pp\to \gamma+jet$ process.}
\label{fig:kinem}
\end{figure}

Fig.~\ref{fig:kinem} shows in the widely used $(x,Q^2)$
kinematic plot (see also \cite{Sti}) what area can be covered for studying the process $q~ g\rrr \gamma +q$.
From this figure (and Tables 1, 2, 5) it becomes clear that even at low LHC luminosity
it would be possible to study the gluon distribution with a good statistics of \gpj events
in the region of small $x$ at values of $Q^2$ that are about 2 orders of magnitude higher than
those reached at HERA now.
It is worth emphasizing that an extension of experimentally reachable 
region at LHC to the region of lower values of $Q^2$, overlapping with the area 
covered by HERA, would be also of a big interest.

~\\[7pt]
\noindent
{\bf \large Acknowledgments} \\[7pt]
We are greatly thankful to D.~Denegri who stimulated us to study the physics of
\gpj processes, permanent support and fruitful suggestions.
It is a pleasure for us to express our recognition for helpful discussions to P.~Aurenche,
M.~Dittmar, M.~Fontannaz, J.Ph.~Guillet, M.L.~Mangano, E.~Pilon,
H.~Rohringer, S.~Tapprogge, H.~Weerts and J.~Womersley.


\newpage

\onecolumn

\noindent
{\bf \large List of figures}\\[2pt]
1. The contributions of various events types to the total number of events as a function of $\Pttg$
presented for three cases: {\bf(a)}~ No separation efficiency is taken into account, 
{\bf(b)}~ $\epsilon^{\,\gamma/mes}$ separation efficiencies are taken into account 
and {\bf(c)}~ $\epsilon^{\,\gamma/mes}$ and $\epsilon^{\,q/g}$ separation
efficiencies  are taken into account.\\
2. LHC  $(x,Q^2)$ kinematic region for $pp\to \gamma+jet$ process.\\[-6mm]

\noindent
{\bf \large List of tables}\\[2pt]
1. Numbers of all events in different $Q^2$ and $x$ intervals for 
$L_{int}=10~fb^{-1}$. \\
2. Numbers of all events in different $Q^2$ and $x$ intervals for 
$L_{int}=10~fb^{-1}$. $\epsilon^{\,\gamma/mes}$ separation efficiencies are taken into account.\\
3. Relative contribution (in per cents) of different QCD subprocesses into
the ``$\gamma\!-\!brem$'' events production.\\
4. Relative contribution (in per cents) of different QCD subprocesses into
the ``$\gamma\!-\!mes$'' events production. \\
5. Numbers of all events in different $Q^2$ and $x$ intervals for 
$L_{int}=10~fb^{-1}$. $\epsilon^{\,\gamma/mes}$ and $\epsilon^{\,q/g}$ separation
efficiencies are taken into account.\\[2pt]
{\bf \large Appendix}\\[2pt]
1A. Numbers of ``$\com$'' events in $Q^2$ and $x$ intervals at $L_{int}=10~fb^{-1}$.\\
2A. Numbers of ``$\ann$'' events in $Q^2$ and $x$ intervals at $L_{int}=10~fb^{-1}$.\\
3A. Numbers of ``$\gamma-brem$'' events in $Q^2$ and $x$ intervals at $L_{int}=10~fb^{-1}$.\\
4A. Numbers of ``$\gamma-mes$'' events in $Q^2$ and $x$ intervals at $L_{int}=10~fb^{-1}$.\\
5A. Numbers of ``$\com$'' events in $Q^2$ and $x$ intervals at $L_{int}=10~fb^{-1}$.
$\epsilon^{\,\gamma/mes}$ separation efficiencies are taken into account.\\
6A. Numbers of ``$\ann$'' events in $Q^2$ and $x$ intervals at $L_{int}=10~fb^{-1}$.
$\epsilon^{\,\gamma/mes}$ separation efficiencies are taken into account.\\
7A. Numbers of ``$\gamma-brem$'' events in $Q^2$ and $x$ intervals at $L_{int}=10~fb^{-1}$.
$\epsilon^{\,\gamma/mes}$ separation efficiencies are taken into account.\\
8A. Numbers of ``$\gamma-mes$'' events in $Q^2$ and $x$ intervals at $L_{int}=10~fb^{-1}$.
$\epsilon^{\,\gamma/mes}$ separation efficiencies are taken into account.\\
9A. Numbers of ``$\com$'' events in $Q^2$ and $x$ intervals at $L_{int}=10~fb^{-1}$.
$\epsilon^{\,\gamma/mes}$ and $\epsilon^{\,q/g}$ separation efficiencies are taken into account.\\
10A. Numbers of ``$\ann$'' events in $Q^2$ and $x$ intervals at $L_{int}=10~fb^{-1}$.
$\epsilon^{\,\gamma/mes}$ and $\epsilon^{\,q/g}$ separation efficiencies are taken into account.\\
11A. Numbers of ``$\gamma-brem$'' events in $Q^2$ and $x$ intervals at $L_{int}=10~fb^{-1}$.
$\epsilon^{\,\gamma/mes}$ and $\epsilon^{\,q/g}$ separation efficiencies are taken into account.\\
12A. Numbers of ``$\gamma-mes$'' events in $Q^2$ and $x$ intervals at $L_{int}=10~fb^{-1}$.
$\epsilon^{\,\gamma/mes}$ and $\epsilon^{\,q/g}$ separation efficiencies are taken into account.\\


~~
\newpage

\begin{table}[htbp]
\begin{center}
\caption{Numbers of all events (divided by $10^3$) in $Q^2$ and $x$ intervals for 
$L_{int}=10~fb^{-1}$.} 
\label{tab:S+B_0}
\vskip0.1cm
\begin{tabular}{|lc|r|r|r|r|>{\columncolor[gray]{\coltab}}r|c|} \hline
 & $Q^2$ &\multicolumn{4}{c|}{ \hspace{-0.9cm} $x$ values of a parton} &All $x$ 
&$\Pt^{\gamma}$   \\\cline{3-7}
 & $(GeV/c)^2$ & $10^{-4}$--$10^{-3}$ & $10^{-3}$--$10^{-2}$ &$10^{-2}$--
$10^{-1}$ & $10^{-1}$--$10^{0}$ & $10^{-4}$--$10^{0}$&$(GeV/c)$     \\\hline
&\hmm\hmm 1600-2500\hmm  & 1393.6  &4301.1  &4506.8  & 481.4  &10682.9    & 40--50\\\hline
&\hmm\hmm 2500-5000\hmm  &  561.1  &2931.0  &3174.7  & 430.4  & 7097.2    & 50--71\\\hline
&\hmm\hmm 5000-10000\hmm &   61.7  & 665.6  & 769.6  & 196.1  & 1693.0    & 71--100\\\hline
&\hmm\hmm 10000-20000\hmm&    3.6  & 150.3  & 178.4  &  81.7  &  414.0    & 100--141\\\hline
&\hmm\hmm 20000-40000\hmm&    0.0  &  29.9  &  40.9  &  25.2  &   96.0    & 141--200 \\\hline
&\hmm\hmm 40000-80000\hmm&    0.0  &   5.7  &  10.7  &   7.8  &   24.2    & 200--283  \\\hline
\multicolumn{6}{c|}{}&{\bf 20~007.3 }\\\cline{7-7}
\end{tabular}
\vskip1.1cm
\caption{Numbers of all events (divided by $10^3$) in $Q^2$ and $x$ intervals for 
$L_{int}=10~fb^{-1}$. $\epsilon^{\,\gamma/mes}$ separation efficiencies are taken into account.}
\label{tab:S+B_1}
\vskip0.2cm
\begin{tabular}{|rr|r|r|r|r|>{\columncolor[gray]{\coltab}}r|c|} \hline
 & $Q^2$ &\multicolumn{4}{c|}{ \hspace{-0.9cm} $x$ values of a parton} &All $x$ 
&$\Pt^{\gamma}$   \\\cline{3-7}
 & $(GeV/c)^2$ & $10^{-4}$--$10^{-3}$ & $10^{-3}$--$10^{-2}$ &$10^{-2}$--
$10^{-1}$ & $10^{-1}$--$10^{0}$ & $10^{-4}$--$10^{0}$&$(GeV/c)$     \\\hline
&\hmm\hmm 1600-2500\hmm  & 1214.6  &3073.1  &3433.1  & 394.5  &8115.4  & 40--50\\\hline
&\hmm\hmm 2500-5000\hmm  &  502.8  &2220.7  &2478.2  & 364.0  &5565.8  & 50--71\\\hline
&\hmm\hmm 5000-10000\hmm &   54.1  & 532.8  & 587.8  & 168.7  &1343.7  & 71--100\\\hline
&\hmm\hmm 10000-20000\hmm&    3.2  & 124.4  & 134.6  &  70.6  & 333.1  & 100--141\\\hline
&\hmm\hmm 20000-40000\hmm&    0.0  &  25.3  &  30.1  &  21.8  &  77.3  & 141--200 \\\hline
&\hmm\hmm 40000-80000\hmm&    0.0  &   4.9  &   7.9  &   6.6  &  19.4  & 200--283  \\\hline
\multicolumn{6}{c|}{}&{\bf 15~454.7 }\\\cline{7-7}
\end{tabular}
\end{center}
\end{table}
\begin{table}[h]
\begin{center}
\vskip20mm
\caption{Relative contribution (in per cents) of main QCD subprocesses into
the ``$\gamma\!-\!brem$'' events production.}
\vskip0.2cm
\begin{tabular}{|c||c|c|c|c|}                  \hline 
\label{tab:bg_or_gr}
$\Pttg$& \multicolumn{4}{c|}{fundamental QCD subprocess} \\\cline{2-5}
 \Gvc & $qg\to qg$ & $qq\to qq$ & $gg\to q\bar{q}$& $gg\to gg$  
\\\hline \hline
 40--71   &  70.6$\pm$ 8.7 & 21.1$\pm$ 3.8 &  5.1$\pm$ 1.6 &  2.6$\pm$ 1.0 \\\hline 
 71--141  &  67.5$\pm$ 7.3 & 23.6$\pm$ 3.5 &  4.2$\pm$ 1.2 &  2.6$\pm$ 0.9  \\\hline 
141--283  &  58.7$\pm$ 9.0 & 30.7$\pm$ 5.7 &  1.8$\pm$ 1.0 &   ---   \\\hline  
\end{tabular}
\vskip20mm
\caption{Relative contribution (in per cents) of main QCD subprocesses into
the ``$\gamma\!-\!mes$'' events production.}
\vskip.2cm
\begin{tabular}{|c||c|c|c|c|}                  \hline 
\label{tab:bg_or_ms}
$\Pttg$& \multicolumn{4}{c|}{fundamental QCD subprocess} \\\cline{2-5}
 \Gvc & $qg\to qg$ & $qq\to qq$ & $gg\to q\bar{q}$& $gg\to gg$ 
\\\hline \hline
 40--71   &  65.2$\pm$ 9.9 & 20.1$\pm$ 4.5 &  7.1$\pm$ 2.5 &  7.2$\pm$ 2.3 \\\hline 
 71--141  &  63.7$\pm$11.6 & 23.0$\pm$ 5.2 &  7.2$\pm$ 2.6 &  4.4$\pm$ 1.4 \\\hline 
141--283  &  57.7$\pm$26.2 & 23.1$\pm$13.9 &  7.7$\pm$ 6.9 &  3.8$\pm$ 4.6 \\\hline 
\end{tabular}
\end{center}
\end{table}

\begin{table}[h]
\begin{center}
\vskip1.1cm
\caption{Numbers of all events (divided by $10^3$) in $Q^2$ and $x$ intervals for 
$L_{int}=10~fb^{-1}$. $\epsilon^{\,\gamma/mes}$ and $\epsilon^{\,q/g}$ separation
efficiencies are taken into account.}
\label{tab:S+B_2}
\vskip0.2cm
\begin{tabular}{|rr|r|r|r|r|>{\columncolor[gray]{\coltab}}r|c|} \hline
 & $Q^2$ &\multicolumn{4}{c|}{ \hspace{-0.9cm} $x$ values of a parton} &All $x$ 
&$\Pt^{\gamma}$   \\\cline{3-7}
 & $(GeV/c)^2$ & $10^{-4}$--$10^{-3}$ & $10^{-3}$--$10^{-2}$ &$10^{-2}$--
$10^{-1}$ & $10^{-1}$--$10^{0}$ & $10^{-4}$--$10^{0}$&$(GeV/c)$     \\\hline
&\hmm\hmm 1600-2500\hmm  &   721.3  & 1858.7  & 2052.9  &  217.6  & 4850.5  & 40--50\\\hline
&\hmm\hmm 2500-5000\hmm  &   302.3  & 1314.1  & 1449.4  &  206.2  & 3271.9  & 50--71\\\hline
&\hmm\hmm 5000-10000\hmm &    31.5  &  320.0  &  350.0  &   99.9  &  801.5  & 71--100\\\hline
&\hmm\hmm 10000-20000\hmm&     1.9  &   74.4  &   81.1  &   41.8  &  199.1  & 100--141\\\hline
&\hmm\hmm 20000-40000\hmm&     0.0  &   14.9  &   18.2  &   12.6  &   45.6  & 141--200 \\\hline
&\hmm\hmm 40000-80000\hmm&     0.0  &    2.9  &    4.5  &    3.8  &   11.2  & 200--283  \\\hline
\multicolumn{6}{c|}{}&{\bf 9~179.8 }\\\cline{7-7}
\end{tabular}
\end{center}
\end{table}

\newpage

\setcounter{table}{0}
\begin{table}[h]
\label{tab:B29_0}
~\\[-5mm]
{\large \bf Appendix}\\[-12pt]
\begin{center}

\def\baselinestretch{1.1}

\small
{{\bf Table 1A}. Numbers of ``$\com$'' events (divided by $10^3$) in $Q^2$ and $x$ 
intervals at $L_{int}=10~fb^{-1}$.}
\vskip0.1cm
\begin{tabular}{|lc|r|r|r|r|>{\columncolor[gray]{\coltab}}r|}  \hline
 & $Q^2$ &\multicolumn{4}{c|}{ \hspace{-0.9cm} $x$ values of a parton} &All $x$   
\\\cline{3-7}
 & $(GeV/c)^2$ & $10^{-4}$--$10^{-3}$ & $10^{-3}$--$10^{-2}$ &$10^{-2}$--
$10^{-1}$ & $10^{-1}$--$10^{0}$ & $10^{-4}$--$10^{0}$     \\\hline
&\hmm\hmm 1600-2500\hmm  &  1040.3  &3128.7  &3202.5  & 275.6  &7647.1  \\\hline
&\hmm\hmm 2500-5000\hmm  &   451.2  &2185.8  &2326.8  & 280.8  &5244.6  \\\hline
&\hmm\hmm 5000-10000\hmm &    45.4  & 545.5  & 611.8  & 151.6  &1354.4  \\\hline
&\hmm\hmm 10000-20000\hmm&     2.9  & 125.5  & 151.1  &  66.7  & 346.2  \\\hline
&\hmm\hmm 20000-40000\hmm&       0  &  24.6  &  35.2  &  19.9  &  79.6  \\\hline
&\hmm\hmm 40000-80000\hmm&       0  &   4.7  &   8.5  &   6.2  &  19.4  \\\hline
\end{tabular}
\vskip0.7cm
{{\bf Table 2A}. Numbers of ``$\ann$'' events (divided by $10^3$) 
in $Q^2$ and $x$ intervals at $L_{int}=10~fb^{-1}$.}
\label{tab:B14_0}
\vskip0.1cm
\begin{tabular}{|lc|r|r|r|r|>{\columncolor[gray]{\coltab}}r|}  \hline
 & $Q^2$ &\multicolumn{4}{c|}{ \hspace{-0.9cm} $x$ values of a parton} &All $x$  
\\\cline{3-7}
 & $(GeV/c)^2$ & $10^{-4}$--$10^{-3}$ & $10^{-3}$--$10^{-2}$ &$10^{-2}$--
$10^{-1}$ & $10^{-1}$--$10^{0}$ & $10^{-4}$--$10^{0}$     \\\hline
&\hmm\hmm 1600-2500\hmm  & 120.3  & 190.2  & 236.8  &  50.5  & 597.8 \\\hline
&\hmm\hmm 2500-5000\hmm  &  43.1  & 239.7  & 250.1  &  35.3  & 568.2 \\\hline
&\hmm\hmm 5000-10000\hmm &   7.7  &  60.5  &  69.0  &  20.5  & 157.7 \\\hline
&\hmm\hmm 10000-20000\hmm&   0.7  &  16.9  &  15.9  &  10.3  &  43.8 \\\hline
&\hmm\hmm 20000-40000\hmm&     0  &   4.2  &   4.4  &   4.2  &  12.8 \\\hline
&\hmm\hmm 40000-80000\hmm&     0  &   0.9  &   1.8  &   1.4  &   4.1  \\\hline
\end{tabular}
\vskip0.7cm
{{\bf Table 3A}. Numbers of ``$\gamma\!-\!brem$'' events (divided by $10^3$)
in $Q^2$ and $x$ intervals at $L_{int}=10~fb^{-1}$.}
\small
\label{tab:g*_0}
\vskip0.1cm
\begin{tabular}{|lc|r|r|r|r|>{\columncolor[gray]{\coltab}}r|}  \hline
 & $Q^2$ &\multicolumn{4}{c|}{ \hspace{-0.9cm} $x$ values of a parton} &All $x$   \\\cline{3-7}
 & $(GeV/c)^2$ & $10^{-4}$--$10^{-3}$ & $10^{-3}$--$10^{-2}$ &$10^{-2}$--
$10^{-1}$ & $10^{-1}$--$10^{0}$ & $10^{-4}$--$10^{0}$     \\\hline
&\hmm\hmm 1600-2500\hmm  &  143.6  & 508.5  & 578.3  & 104.8  &1335.3  \\\hline
&\hmm\hmm 2500-5000\hmm  &   51.3  & 328.2  & 432.1  &  94.8  & 906.5  \\\hline
&\hmm\hmm 5000-10000\hmm &    4.3  &  42.0  &  59.0  &  13.7  & 119.0  \\\hline
&\hmm\hmm 10000-20000\hmm&      0  &   5.2  &   9.2  &   2.8  &  17.2  \\\hline
&\hmm\hmm 20000-40000\hmm&      0  &   0.9  &   0.9  &   1.0  &   2.8  \\\hline
&\hmm\hmm 40000-80000\hmm&      0  &   0.1  &   0.4  &   0.2  &   0.7  \\\hline
\end{tabular}
\vskip0.7cm
{{\bf Table 4A}. Numbers of ``$\gamma\!-\!mes$'' events (divided by $10^3$) 
in $Q^2$ and $x$ intervals at $L_{int}=10~fb^{-1}$.}
\small
\label{tab:pi0_0}
\vskip0.1cm
\begin{tabular}{|lc|r|r|r|r|>{\columncolor[gray]{\coltab}}r|}  \hline
 & $Q^2$ &\multicolumn{4}{c|}{ \hspace{-0.9cm} $x$ values of a parton} &All $x$    \\\cline{3-7}
 & $(GeV/c)^2$ & $10^{-4}$--$10^{-3}$ & $10^{-3}$--$10^{-2}$ &$10^{-2}$--
$10^{-1}$ & $10^{-1}$--$10^{0}$ & $10^{-4}$--$10^{0}$     \\\hline
&\hmm\hmm 1600-2500\hmm  &  89.3  & 473.6  & 489.1  &  50.5  &1102.4  \\\hline
&\hmm\hmm 2500-5000\hmm  &  15.5  & 177.3  & 165.6  &  19.4  & 377.7  \\\hline
&\hmm\hmm 5000-10000\hmm &   4.3  &  17.6  &  29.5  &  10.3  &  61.6  \\\hline
&\hmm\hmm 10000-20000\hmm&     0  &   2.6  &   2.2  &   1.9  &   6.7  \\\hline
&\hmm\hmm 20000-40000\hmm&     0  &   0.2  &   0.4  &   0.2  &   0.8  \\\hline
&\hmm\hmm 40000-80000\hmm&     0  &     0  &  0.02  &  0.01  &  0.03  \\\hline
\end{tabular}
\end{center}
\end{table}

\def\baselinestretch{1.1}

\begin{table}[h]
\begin{center}
\small
{{\bf Table 5A}. Numbers of ``$\com$'' events (divided by $10^3$)
in $Q^2$ and $x$ intervals at $L_{int}=10~fb^{-1}$.
$\epsilon^{\,\gamma/mes}$ separation efficiencies are taken into account.}
\label{tab:B29_1}
\vskip0.1cm
\begin{tabular}{|lc|r|r|r|r|>{\columncolor[gray]{\coltab}}r|}  \hline
 & $Q^2$ &\multicolumn{4}{c|}{ \hspace{-0.9cm} $x$ values of a parton} &All $x$  \\\cline{3-7}
 & $(GeV/c)^2$ & $10^{-4}$--$10^{-3}$ & $10^{-3}$--$10^{-2}$ &$10^{-2}$--
$10^{-1}$ & $10^{-1}$--$10^{0}$ & $10^{-4}$--$10^{0}$     \\\hline
&\hmm\hmm 1600-2500\hmm  &  945.0  &2387.4  &2608.6  & 246.7  &6187.8  \\\hline
&\hmm\hmm 2500-5000\hmm  &  410.5  &1723.1  &1865.0  & 253.0  &4251.7  \\\hline
&\hmm\hmm 5000-10000\hmm &   41.3  & 443.2  & 475.5  & 134.6  &1094.6  \\\hline
&\hmm\hmm 10000-20000\hmm&    2.6  & 105.0  & 114.5  &  58.7  & 281.0  \\\hline
&\hmm\hmm 20000-40000\hmm&      0  &  20.9  &  25.9  &  17.3  &  64.2  \\\hline
&\hmm\hmm 40000-80000\hmm&      0  &   4.0  &   6.3  &   5.3  &  15.7  \\\hline
\end{tabular}
\vskip0.7cm
\hspace*{-1.5mm} 
{{\bf Table 6A}. Numbers of ``$\ann$'' events (divided by $10^3$)
in $Q^2$ and $x$ intervals at $L_{int}=10~fb^{-1}$.
$\epsilon^{\,\gamma/mes}$ separation efficiencies are taken into account.}
\label{tab:B14_1}
\vskip0.1cm
\begin{tabular}{|lc|r|r|r|r|>{\columncolor[gray]{\coltab}}r|}  \hline
 & $Q^2$ &\multicolumn{4}{c|}{ \hspace{-0.9cm} $x$ values of a parton} &All $x$   \\\cline{3-7}
 & $(GeV/c)^2$ & $10^{-4}$--$10^{-3}$ & $10^{-3}$--$10^{-2}$ &$10^{-2}$--
$10^{-1}$ & $10^{-1}$--$10^{0}$ & $10^{-4}$--$10^{0}$     \\\hline
&\hmm\hmm 1600-2500\hmm  & 109.5  & 142.9  & 192.6  & 451.0  & 490.2 \\\hline
&\hmm\hmm 2500-5000\hmm  &  39.2  & 185.1  & 196.8  &  29.7  & 451.0 \\\hline
&\hmm\hmm 5000-10000\hmm &   7.0  &  48.7  &  54.1  &  17.9  & 127.8 \\\hline
&\hmm\hmm 10000-20000\hmm&   0.6  &  13.8  &  12.1  &   8.8  &  35.4 \\\hline
&\hmm\hmm 20000-40000\hmm&     0  &   3.5  &   3.2  &   3.5  &  10.2 \\\hline
&\hmm\hmm 40000-80000\hmm&     0  &   0.7  &   1.3  &   1.1  &   3.2 \\\hline
\end{tabular}
\vskip0.7cm
{{\bf Table 7A}. Numbers of ``$\gamma\!-\!brem$'' events (divided by $10^3$)
in $Q^2$ and $x$ intervals at 
$L_{int}=10~fb^{-1}$. $\epsilon^{\,\gamma/mes}$ separation efficiencies are taken into account.}
\small
\label{tab:g*_1}
\vskip0.1cm
\begin{tabular}{|lc|r|r|r|r|>{\columncolor[gray]{\coltab}}r|}  \hline
 & $Q^2$ &\multicolumn{4}{c|}{ \hspace{-0.9cm} $x$ values of a parton} &All $x$    \\\cline{3-7}
 & $(GeV/c)^2$ & $10^{-4}$--$10^{-3}$ & $10^{-3}$--$10^{-2}$ &$10^{-2}$--
$10^{-1}$ & $10^{-1}$--$10^{0}$ & $10^{-4}$--$10^{0}$     \\\hline
&\hmm\hmm 1600-2500\hmm  &  129.9  & 394.3  & 476.6  &  87.2  &1088.0  \\\hline
&\hmm\hmm 2500-5000\hmm  &   46.7  & 258.5  & 359.8  &  74.8  & 739.8  \\\hline
&\hmm\hmm 5000-10000\hmm &    3.9  &  34.0  &  47.1  &  11.7  &  96.6  \\\hline
&\hmm\hmm 10000-20000\hmm&      0  &   4.4  &   7.1  &   2.4  &  13.9  \\\hline
&\hmm\hmm 20000-40000\hmm&      0  &   0.8  &   0.7  &   0.9  &   2.4  \\\hline
&\hmm\hmm 40000-80000\hmm&      0  &   0.1  &   0.3  &   0.2  &   0.6  \\\hline
\end{tabular}
\vskip0.7cm
{{\bf Table 8A}. Numbers of ``$\gamma\!-\!mes$'' events (divided by $10^3$)
in $Q^2$ and $x$ intervals at 
$L_{int}=10~fb^{-1}$. $\epsilon^{\,\gamma/mes}$ separation efficiencies are taken into account.}
\small
\label{tab:pi0_1}
\vskip0.1cm
\begin{tabular}{|lc|r|r|r|r|>{\columncolor[gray]{\coltab}}r|}  \hline
 & $Q^2$ &\multicolumn{4}{c|}{ \hspace{-0.9cm} $x$ values of a parton} &All $x$   
\\\cline{3-7}
 & $(GeV/c)^2$ & $10^{-4}$--$10^{-3}$ & $10^{-3}$--$10^{-2}$ &$10^{-2}$--
$10^{-1}$ & $10^{-1}$--$10^{0}$ & $10^{-4}$--$10^{0}$     \\\hline
&\hmm\hmm 1600-2500\hmm  &  30.2  & 148.5  & 155.2  &  15.4  & 349.4  \\\hline
&\hmm\hmm 2500-5000\hmm  &   6.4  &  53.9  &  56.6  &   6.4  & 123.3  \\\hline
&\hmm\hmm 5000-10000\hmm &   1.9  &   6.9  &  11.2  &   4.6  &  24.6  \\\hline
&\hmm\hmm 10000-20000\hmm&     0  &   1.1  &   0.9  &   0.8  &   2.8  \\\hline
&\hmm\hmm 20000-40000\hmm&     0  &   0.1  &   0.3  &   0.1  &   0.5  \\\hline
&\hmm\hmm 40000-80000\hmm&     0  &     0  &  0.01  &  0.01  &  0.02  \\\hline
\end{tabular}
\end{center}
\end{table}

\def\baselinestretch{1.1}

\begin{table}[h]
\begin{center}
~\\[-5mm]
\small
{{\bf Table 9A}. Numbers of ``$\com$'' events (divided by $10^3$)
in $Q^2$ and $x$ intervals at $L_{int}=10~fb^{-1}$.
$\epsilon^{\,\gamma/mes}$ and $\epsilon^{\,q/g}$ separation efficiencies  are taken into account.}
\label{tab:B29_2}
\vskip0.1cm
\begin{tabular}{|lc|r|r|r|r|>{\columncolor[gray]{\coltab}}r|}  \hline
 & $Q^2$ &\multicolumn{4}{c|}{ \hspace{-0.9cm} $x$ values of a parton} &All $x$   \\\cline{3-7}
 & $(GeV/c)^2$ & $10^{-4}$--$10^{-3}$ & $10^{-3}$--$10^{-2}$ &$10^{-2}$--
$10^{-1}$ & $10^{-1}$--$10^{0}$ & $10^{-4}$--$10^{0}$     \\\hline
&\hmm\hmm 1600-2500\hmm  &  623.7  &1575.7  &1721.7  & 162.8  &4084.0  \\\hline
&\hmm\hmm 2500-5000\hmm  &  271.0  &1137.3  &1230.9  & 167.0  &2806.2  \\\hline
&\hmm\hmm 5000-10000\hmm &   27.3  & 292.5  & 313.8  &  88.9  & 722.5  \\\hline
&\hmm\hmm 10000-20000\hmm&    1.7  &  69.4  &  75.6  &  38.7  & 185.5  \\\hline
&\hmm\hmm 20000-40000\hmm&    0.0  &  13.8  &  17.1  &  11.5  &  42.4  \\\hline
&\hmm\hmm 40000-80000\hmm&    0.0  &   2.7  &   4.2  &   3.5  &  10.4  \\\hline
\end{tabular}
\vskip0.7cm
{{\bf Table 10A}. Numbers of ``$\ann$'' events (divided by $10^3$)
in $Q^2$ and $x$ intervals at $L_{int}=10~fb^{-1}$.
$\epsilon^{\,\gamma/mes}$ and $\epsilon^{\,q/g}$ separation efficiencies  are taken into account.}
\label{tab:B14_2}
\vskip0.1cm
\begin{tabular}{|lc|r|r|r|r|>{\columncolor[gray]{\coltab}}r|}  \hline
 & $Q^2$ &\multicolumn{4}{c|}{ \hspace{-0.9cm} $x$ values of a parton} &All $x$    \\\cline{3-7}
 & $(GeV/c)^2$ & $10^{-4}$--$10^{-3}$ & $10^{-3}$--$10^{-2}$ &$10^{-2}$--
$10^{-1}$ & $10^{-1}$--$10^{0}$ & $10^{-4}$--$10^{0}$     \\\hline
&\hmm\hmm 1600-2500\hmm  &  29.1  &  37.8  &  51.0  &  12.0  & 129.9  \\\hline
&\hmm\hmm 2500-5000\hmm  &   9.9  &  46.2  &  48.9  &   7.5  & 112.4  \\\hline
&\hmm\hmm 5000-10000\hmm &   1.5  &  10.7  &  11.9  &   3.9  &  27.9  \\\hline
&\hmm\hmm 10000-20000\hmm&   0.1  &   2.6  &   2.3  &   1.7  &   6.7  \\\hline
&\hmm\hmm 20000-40000\hmm&   0.0  &   0.7  &   0.6  &   0.7  &   1.9  \\\hline
&\hmm\hmm 40000-80000\hmm&   0.0  &   0.1  &   0.2  &   0.2  &   0.6  \\\hline
\end{tabular}
\vskip0.7cm
{{\bf Table 11A}. 
Numbers of ``$\gamma\!-\!brem$'' events (divided by $10^3$)
in $Q^2$ and $x$ intervals at $L_{int}=10~fb^{-1}$.
$\epsilon^{\,\gamma/mes}$ and $\epsilon^{\,q/g}$ separation efficiencies  are taken into account.}
\small
\label{tab:g*_2}
\vskip0.1cm
\begin{tabular}{|lc|r|r|r|r|>{\columncolor[gray]{\coltab}}r|}  \hline
 & $Q^2$ &\multicolumn{4}{c|}{ \hspace{-0.9cm} $x$ values of a parton} &All $x$    \\\cline{3-7}
 & $(GeV/c)^2$ & $10^{-4}$--$10^{-3}$ & $10^{-3}$--$10^{-2}$ &$10^{-2}$--
$10^{-1}$ & $10^{-1}$--$10^{0}$ & $10^{-4}$--$10^{0}$     \\\hline
&\hmm\hmm 1600-2500\hmm  &   48.5  &  147.1  &  177.8  &   32.5  &  406.0  \\\hline
&\hmm\hmm 2500-5000\hmm  &   17.2  &   95.0  &  132.2  &   27.5  &  272.0  \\\hline
&\hmm\hmm 5000-10000\hmm &    1.4  &   12.2  &   17.0  &    4.2  &   34.8  \\\hline
&\hmm\hmm 10000-20000\hmm&    0.0  &    1.6  &    2.6  &    0.9  &    5.1  \\\hline
&\hmm\hmm 20000-40000\hmm&    0.0  &    0.3  &    0.3  &    0.3  &    0.9  \\\hline
&\hmm\hmm 40000-80000\hmm&    0.0  &    0.0  &    0.1  &    0.1  &    0.2  \\\hline
\end{tabular}
\vskip0.7cm
{{\bf Table 12A}. Numbers of ``$\gamma\!-\!mes$'' events (divided by $10^3$)
in $Q^2$ and $x$ intervals at $L_{int}=10~fb^{-1}$.
$\epsilon^{\,\gamma/mes}$ and $\epsilon^{\,q/g}$ separation efficiencies  are taken into account.}
\small
\label{tab:pi0_2}
\vskip0.1cm
\begin{tabular}{|lc|r|r|r|r|>{\columncolor[gray]{\coltab}}r|}  \hline
 & $Q^2$ &\multicolumn{4}{c|}{ \hspace{-0.9cm} $x$ values of a parton} &All $x$    \\\cline{3-7}
 & $(GeV/c)^2$ & $10^{-4}$--$10^{-3}$ & $10^{-3}$--$10^{-2}$ &$10^{-2}$--
$10^{-1}$ & $10^{-1}$--$10^{0}$ & $10^{-4}$--$10^{0}$     \\\hline
&\hmm\hmm 1600-2500\hmm  &  19.9  &   98.0  &  102.5  &   10.2  &  230.6 \\\hline
&\hmm\hmm 2500-5000\hmm  &   4.2  &   35.6  &   37.4  &    4.2  &   81.4 \\\hline
&\hmm\hmm 5000-10000\hmm &   1.3  &    4.6  &    7.4  &    3.0  &   16.2 \\\hline
&\hmm\hmm 10000-20000\hmm&   0.0  &    0.8  &    0.6  &    0.5  &    1.9 \\\hline
&\hmm\hmm 20000-40000\hmm&   0.0  &    0.1  &    0.2  &    0.1  &    0.3 \\\hline
&\hmm\hmm 40000-80000\hmm&   0.0  &    0.0  &   0.01  &   0.01  &   0.02 \\\hline
\end{tabular}
\end{center}
\end{table}

\end{document}